\documentclass[twocolumn,showpacs,preprintnumbers,amsmath,amssymb]{revtex4}
\usepackage{graphicx}
\usepackage{dcolumn}

\bibliography{2}

\begin{document}
{\bf Comment on ``Total Negative Refraction in Crystals for
Ballistic Electrons and Light''} \vspace{3mm}

 The effect of negative refraction considered in Ref.\cite{1}
resembles the analogous effect in left-handed materials (LHM)
although the physics is different. In LHM the negative refraction
is caused by a negative group velocity, whereas in the case
considered in Ref.\cite{1} one deals with the special properties
of wave propagation in uniaxial anisotropic media (UAM).

The dispersion relation (DR) for an electromagnetic plane wave in
UAM has the form\cite{2}:
$\omega^2/c^2=k_\parallel^2/\varepsilon_\perp+k_\perp^2/\varepsilon_\parallel$,
where $k_\parallel$ and $k_\perp$ are the components of the wave
vector $\bf{k}$ parallel and perpendicular to the uniaxis, and
$\varepsilon_\parallel$ and $\varepsilon_\perp$ are the
corresponding dielectric tensor components in a principal
coordinate system. It follows from the DR that ${\bf
v}_{ph}\cdot{\bf v}_{g}>0$, whereas in LHM ${\bf v}_{ph}\cdot{\bf
v}_{g}<0$, ${\bf v}_{ph}$ and ${\bf v}_{g}$ being the phase and
group velocities, respectively. It also follows from the DR that
the directions of the phase and group velocities are connected by
the relation
$\tan\varphi_g=(\varepsilon_\perp/\varepsilon_\parallel)\tan\varphi
$, where $\varphi $ and $\varphi_g$ are the angles between the
anisotropy direction ${\bf n}$ and the vectors ${\bf v}_{ph}$ (or
${\bf k}$) and ${\bf v}_g$, respectively. The relation between the
directions of ${\bf v}_g$ and ${\bf v}_{ph}$ can be determined
graphically. As an example we will consider the case of
$\varepsilon_\perp/\varepsilon_\parallel>1$. The end of the vector
$\bf{k}$ lies on an ellipse in the $(k_\perp,\,k_\parallel)$ plane
(Fig.1). Let us introduce another ellipse, which is obtained from
the first one by stretching it along the $k_\perp$ axis
$\varepsilon_\perp/\varepsilon_\parallel$ times. The vector ${\bf
k}_g$, whose end lies on the second (bigger) ellipse and has the
same projection on the $k_\parallel$ axis as the vector $\bf k$,
is parallel to ${\bf v}_g$. In such a way one can obtain the
direction of the group velocity using the direction of the vector
$\bf k$ and vice versa. For simplicity we consider here and below
2D geometry.
\begin{figure}[htb]
\centering \scalebox{0.195}{\includegraphics{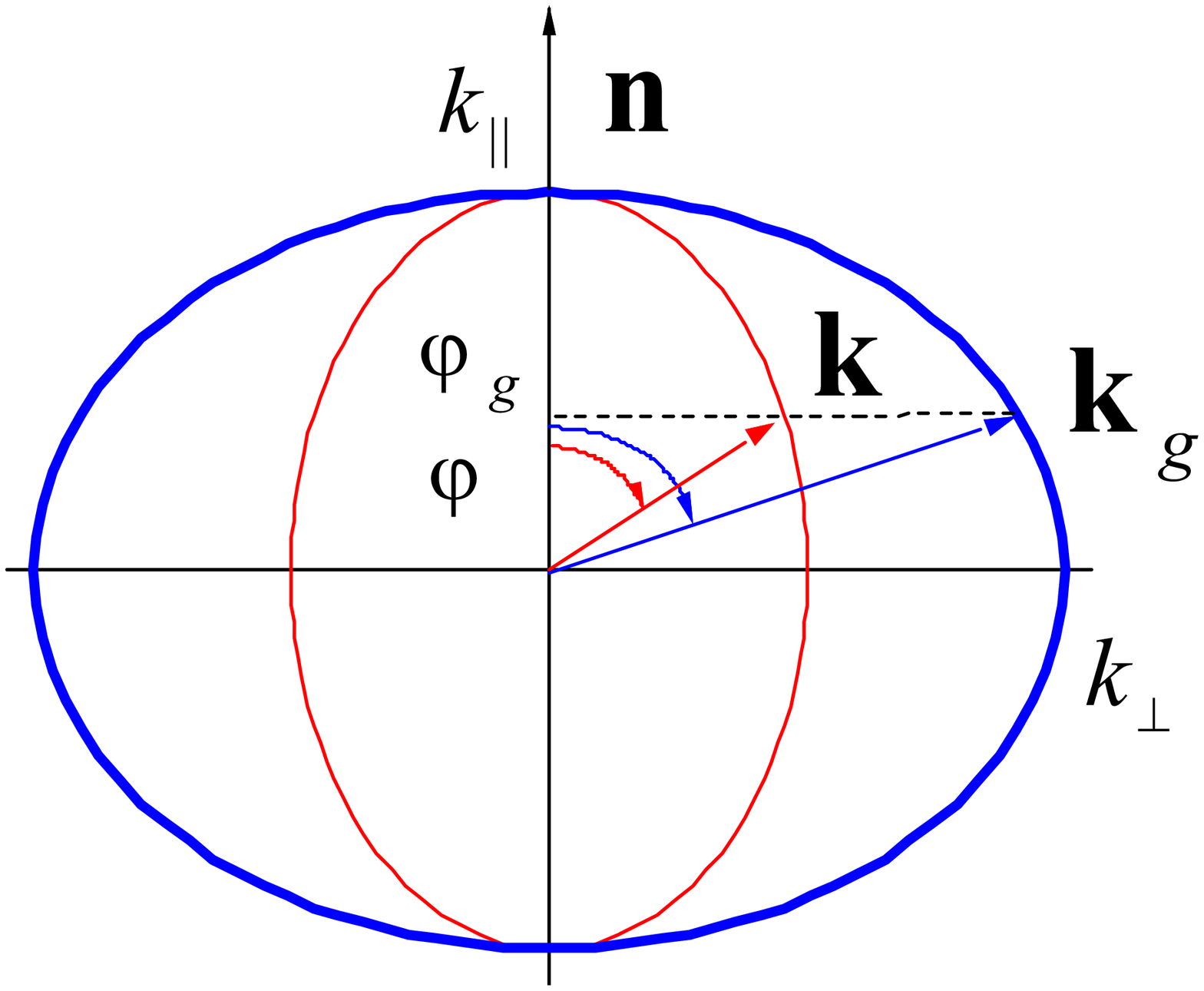}}
 \caption{Relation between  $\bf k$ and ${\bf k}_g$.}
\end{figure}

Using this method it is easy to consider the refraction of rays on
the interface between two identical but arbitrarily oriented UAM
(Fig.2). Keeping in mind that the direction of the ray coincides
with the group velocity direction, one can determine the sectors
$A_1^\prime B_1^\prime$ and $B_1^\prime C_1^\prime$ of the small
ellipse, which correspond to rays with negative (sector $A_1 B_1$
of the big ellipse) and positive (sector $B_1 C_1$) angles of
incidence. The end of the wave vector of the refracted ray lies on
a similar but differently oriented ellipse and has the same
projection on the interface plane as the wave vector of the
incident ray. In this way one can determine the wave vector of the
refracted ray and, using the graphical method, the refracted ray
direction. Dividing the right ellipses to regions corresponding to
negative (big ellipse sector $A_2B_2$ and small ellipse sector
$A_2^\prime B_2^\prime$) and positive (sectors $B_2C_2$ and
$B_2^\prime C_2^\prime$) angles of refraction, it is easy to see
that rays with negative angles of incidence lying in the sector
$D_1B_1$, transform to rays with positive angles of refraction
(sector $B_2D_2$).

A general picture of refraction of arbitrary rays is presented in
Fig.3. The incident ray $b$ and the refracted ray $b^\prime$ lie
on the same side of the normal to the interface that seems as
negative refraction, but physically this effect has nothing in
common with negative refraction in LHM.
\begin{figure}[htb]
\centering \scalebox{0.52}{\includegraphics{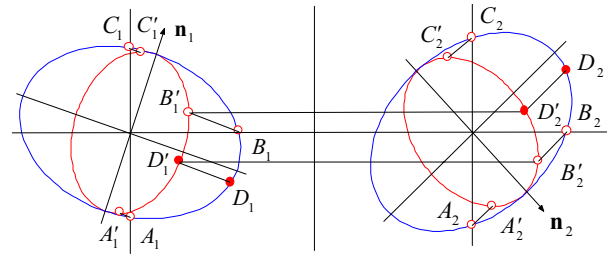}}
 \caption{Determination of regions of positive and negative refraction.}
\end{figure}
\begin{figure}[htb]
\centering \scalebox{0.2}{\includegraphics{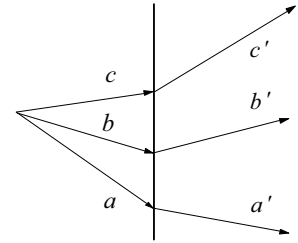}}
 \caption{Refraction of arbitrary rays.}
\end{figure}

Note, that a periodic structure composed of alternate layers with
different permittivity $\varepsilon_1$ and $\varepsilon_2$ behaves
as a UAM with respect to electromagnetic waves whose wavelength is
much larger than the structure period.\\

Yu.P.Bliokh and J.Felsteiner\\
{\it Department of Physics, Technion,
\\ 32000 Haifa, Israel.}

\end{document}